\documentclass[showpacs,amsmath,amssymb,aps,twocolumn]{revtex4-1}

\usepackage[utf8]{inputenc}
\usepackage[english]{babel}
\usepackage{esint}
\usepackage{braket}
\usepackage{dcolumn}
\usepackage{verbatim}

\usepackage{graphicx}
\usepackage{epstopdf} 

\usepackage[dvipsnames]{xcolor}
\usepackage[normalem]{ulem}
\usepackage{hyperref}
\usepackage{slashed}
\usepackage{dsfont}
\usepackage{float}
\usepackage{latexsym}
\usepackage{epsf}

\newcommand{\be}[1]{ \begin{eqnarray} \mbox{$\label{#1}$} }
   
\newcommand{\ee}{\end{eqnarray}}

\newcommand{\eeq}{\end{equation}}

\newcommand\iec {{\it i.e., }}

\newcommand\half{\frac 1 2 }

  \DeclareMathOperator{\Tr}{Tr}
  \DeclareMathOperator{\sgn}{sgn}

\def\a{\alpha}
\def\b{\beta}

\def\k{\kappa}
\def\l{\lambda}

\newcommand{\vk}{\vec{k}}
\newcommand{\vA}{\vec{A}}
\newcommand{\vB}{\vec{B}}

\newcommand{\vq}{\vec{q}}
\newcommand{\sech}{\text{sech}}
\newcommand{\vb}{\vec{b}}
\newcommand{\vsigma}{\vec{\sigma}}
\newcommand{\vgamma}{\vec{\gamma}}

\addto\captionsenglish{}

\begin{document}
\title{Emergent Chern-Simons Interactions in 3+1 Dimensions}

\author{M. St\aa lhammar$^{1,2}$}
\author{D. Rudneva$^2$}
\author{T.H. Hansson$^2$}
\author{F. Wilczek$^{1,2,3,4,5}$}

\affiliation{$^3$Center for Theoretical Physics, MIT, Cambridge, Massachusetts 02139 USA}
\affiliation{$^4$T. D. Lee Institute and Wilczek Quantum Center, \\
Shanghai Jiao Tong University, Shanghai, China}
\affiliation{$^5$Arizona State University, Tempe, Arizona  85287, USA}

\begin{abstract}
Parity-violating superconductors can support a low-dimension local interaction that becomes, upon condensation, a purely spatial Chern-Simons term.  Solutions to the resulting generalized London equations can be obtained from solutions of the ordinary London equations with a complex penetration depth, and suggest several remarkable physical phenomena.  The problem of flux exclusion by a sphere brings in an anapole moment, the problem of current-carrying wires brings in an azimuthal magnetic field, and the problem of vortices brings in currents along the vortices.   We demonstrate that interactions of this kind, together with a conceptually related dimensionally reduced Chern-Simons interaction, can arise from physically plausible microscopic interactions.  
\end{abstract}
                                        
\maketitle                 

\section{Introduction}

The principles of symmetry and locality allow us to survey interactions that are likely to emerge in the description of materials at low energy in a systematic way.  This Ginzburg-Landau or effective field theory approach has proved to be  a fruitful guide to low-energy dynamics, phase transitions, and response to external fields in many applications. In this approach, the focus is on possible interactions represented by local operators of low mass dimension.  
In this paper we shall study terms in the effective Lagrangian, specifically $\vec \b \cdot \vec A\times\vec E - A_0 \vec \beta \cdot \vec B$  and especially $\vec A\cdot\vec B$, that break discrete symmetries. For general reasons related to gauge symmetry it appears that the second of these terms cannot appear in a normal material in its thermodynamic ground state \cite{nazarov86}.  However, as demonstrated below, it is permitted in superconducting states that can support persistent current. 
We will provide examples of microscopic models that illustrate both terms. These terms have noteworthy phenomenological implications, which we will explore further and exemplify below.

Interactions mediated by Lagrangian densities of the Chern-Simons form 
\begin{equation}\label{spatial_CS}
{\cal L}  ~\propto~ \epsilon^{\alpha \beta \gamma} A_\alpha \partial_\beta A_\gamma,
\end{equation}
involving a gauge field $A_\alpha$, and their multi-field and non-Abelian generalizations, have attracted much attention in recent years, mostly in the context of (2+1)-dimensional systems, where they can have a topological character. 
In its most straightforward application, the gauge field $\vA$ is the gauge field of electromagnetism.  Then these terms directly induce, and parametrize, interesting aspects of electromagnetic response that, at a heuristic level, derive from current-field mixing. 

Here we will examine a different appearance of interactions of this kind, in 3+1 dimensions, where we take all the indices to be spatial:
\begin{equation}\label{CSt}
{\cal L}_{\rm CSt} = -\frac \beta 2 \vec A \cdot \vec B.
\end{equation}
Terms of this kind are not relativistically invariant, and they also violate parity (but not time reversal).   That does not forbid their appearance, since many materials, such as those based on crystals lacking an inversion center,  and others described below, violate those symmetries.   More seriously, such terms embody only a limited form of gauge symmetry.  Under a local gauge transformation $A_\alpha \rightarrow A_\alpha + \partial_\alpha \Lambda$, we have
\begin{equation}
\epsilon^{\alpha \beta \gamma} A_\alpha \partial_\beta A_\gamma ~\rightarrow~  \epsilon^{\alpha \beta \gamma} A_\alpha \partial_\beta A_\gamma \, + \, \partial_\alpha (\epsilon^{\alpha \beta \gamma} \Lambda \partial_\beta A_\gamma ),
\end{equation}
so the change in the bulk interaction can be cast into a surface term; but in possible applications the surface term requires careful consideration.  Notably, in the context of the quantum Hall effect it is connected to the existence of edge modes and is canceled through an anomalous surface theory \cite{stone1991edge}.   In addition to the spatial Chern-Simons term Eq.~\eqref{CSt}, we shall also consider the term
\be{CSs}
{\cal L}_{\rm CSs} = \frac{1}{2} \, \bigl( \vec\beta\cdot \vec A\times\vec E -  A_0  \vec \beta \cdot \vec B \bigr).
\ee
 
In a superconductor we can generate a spatial Chern-Simons interaction from a conventional, manifestly gauge invariant interaction by condensation, viz.:
\begin{equation}
{\rm Re} \,  \phi^\dagger i \vec D \phi \cdot \vec B ~\rightarrow~ q v^2 \vec A\cdot \vec B,
\end{equation}
where $\vec D \equiv \vec\nabla - iq \vec A$ is the covariant derivative, $\phi \rightarrow \langle \phi \rangle \equiv v$ through condensation, and where we used the London gauge.  This is similar to how condensation generates a photon mass term $\propto A^2$ from the kinetic energy $\propto \phi^\dagger \vec D^2 \phi$.   We can expect such terms to arise even in $s$-wave superconductors that violate parity symmetry, for example those based on chiral crystals, on organic superconductors subject to chiral selection of the base molecules, generic $s$-wave superconductors incorporating chiral dopants, or non-centrosymmetric superconductors \cite{PhysRevB.70.104521,AGTERBERG200313,Bauer2012NonCentrosymmetricS,PhysRevB.102.184517}. 

Spatial Chern-Simons terms are also known to be directly connected to the chiral magnetic effect (CME), and arise naturally  in various Weyl systems as a consequence of driving or domain wall motion \cite{PhysRevB.102.241401,10.21468/SciPostPhys.10.5.102,PhysRevLett.116.077201,Landsteiner}. The presence of these terms have, however, been debated in the past years, for instance, in Refs~\cite{PhysRevLett.111.027201,  PhysRevB.89.035142} (see, also Refs.~\cite{HOSUR2013857,Landsteiner} for a summary). Care must be taken to distinguish the equilibrium ground state  from driven states, and also to take the correct order of the limits  $\vec{q}\to0$ and $\omega \to 0$ \cite{PhysRevB.88.125105,PhysRevB.91.115203}.   In Sec.~\ref{sect:micro} below, we will further investigate the presence of both $\vec{A}\cdot\vec{B}$, and $\vec\beta\cdot \vec A\times\vec E $ terms in equilibrium Weyl systems.

Heuristically, one identifies quantities $ \vec  {\tilde j }$  that appear in Lagrangian densities of the form $\vec A \cdot \vec {\tilde j }$ as effective currents, since they will  appear as such in the Maxwell equations.  (If $\vec {\tilde j}$ depends explicitly on $\vec A$, slight complications ensue.)   Famously, the London diamagnetic current ${\vec j}_d \propto \vec A$ is characteristic of the superconducting photon mass. 
Following this heuristic, the spatial Chern-Simons term Eq.\,(\ref{CSt}) gives us a current ${\vec j}_{\text{CS}}$ proportional to the magnetic field:
\begin{equation}\label{current_field}
\vec j_{\text{CS}} \propto \vec B.
\end{equation}
This yields unusual, interesting, and potentially important phenomenological consequences, which is the subject of Sect.~\ref{sect:abphen}

From a broader theoretical perspective, a natural term descending from a Lorentz invariant effective action is $ \epsilon^{\mu\nu\sigma\omega} \beta_\mu A_\nu F_{\sigma\omega}$, where $\beta_\mu$ is an axial ({\it i.e.}, unnatural) four-vector. Constant values of $\beta$ violate Lorentz invariance, and a constant $\beta_0$ can be powerfully constrained phenomenologically using astronomical data \cite{PhysRevD.41.1231}. But interactions of the form Eqs.\,(\ref{CSt}) and \eqref{CSs} arise naturally from the canonical axion coupling to electromagnetic fields
\begin{equation}
{\cal L} ~\propto~ a \epsilon^{\alpha\beta\gamma\delta} \partial_{\alpha} A_\beta \partial_{\gamma} A_\delta
\end{equation} 
for the simplest space-time variations of $a$, corresponding to the axion background $a=\beta t$ and $a = \vec\beta\cdot\vec x$, respectively, where the latter describes  an ``axion wind'' background, \iec one that is constant in time but varies linearly in a spatial direction.  Of course, both can occur together.
 
The expression $\vec A \cdot\vec B$ appears in many places in the literature on  magnetohydrodynamics \cite{moffatt69}, where it is used to characterize the magnetic field configurations in the plasmas. In particular, Eq.~\eqref{current_field} describes a  force-free field since the Lorentz force on a current parallel with the magnetic field vanishes \cite{woltjer1958theorem}.  We stress, however, that here we are interested in the response of materials where  $\vec A \cdot\vec B$ is part of the effective action, and thus determines the response to external fields.

\section{Phenomena in Actively Chiral Superconductors}
\label{sect:abphen}

For ease of reference, and in view of their connection with chirality and optical activity, we shall refer to superconductors that incorporate a purely spatial Chern-Simons term ${\cal L}_{\text{CS}}$ as {\it actively chiral\/}  superconductors.  We will work with the Lagrangian density
\begin{equation}\label{L_density}
{\cal L} ~=~  \half E^2 - \half B^2 - \frac{\beta}{2} \vA \cdot  \vB - \frac{\gamma}{2}  A^2.
\end{equation}

\subsection{Plane waves and stability}

From Eq.\,(\ref{L_density}) we derive, after fixing the gauge $A_0 = 0$ and adopting the plane-wave {\it ansatz\/} 
\begin{equation}
\vA ~=~ \vec{\varepsilon} \exp i (\vk \cdot \vec{x} - \omega t)
\end{equation}
the equations of motion
\begin{eqnarray}
\vk \cdot \vec{\varepsilon} ~&=&~ 0, \\
(\omega^2 - k^2 - \gamma) \vec{\varepsilon} \, \mp \, \beta \vec{k}  \times \vec{\varepsilon} ~&=&~ 0.
\end{eqnarray}
The eigen-polarizations are transverse and circular.  Indeed, with
\begin{eqnarray}
\vk ~&=&~ (0, 0, k), \nonumber \\
\vec{\varepsilon} ~&\propto&~  (1, \pm i, 0) ,
\end{eqnarray}
we find the dispersion relations
\begin{equation}
(\omega^2 - k^2 - \gamma) \mp \beta k ~=~ 0.
\end{equation}

The two circular polarizations propagate with different velocities.  This gives rise to optical activity, i.e., rotation of the plane of linear polarization as the (transverse) wave propagates.

For stability in time we require that for real $k$ the $\omega$ that solve the dispersion relation are real.  This gives us the stability condition
\begin{equation}\label{stability}
4 \gamma \geq \beta^2.
\end{equation}
The same condition also ensures the positivity of the energy.  Indeed, since the electric field contribution is manifestly positive, at issue is only the positivity of the magnetic energy
\begin{equation}
{\cal E} ~=~ \frac{1}{2} \, \int B^2 + \beta \vA\cdot \vB + \gamma A^2.
\end{equation}
We can write this as
\begin{align}
{\cal E} = \frac{1}{4} \, \int &\left(1 + \frac{\beta}{2\sqrt \gamma}\right) \left(\vB + \sqrt \gamma \vA\right)^2  \, + \nonumber
\\
&\left(1 - \frac{\beta}{2\sqrt \gamma}\right) \left(\vB - \sqrt \gamma \vA\right)^2 .
\end{align}
When Eq.\,(\ref{stability}) is satisfied the coefficients of these two manifestly positive terms will both be non-negative. 

The stability condition Eq.\,(\ref{stability}) requires, for $\beta \neq 0$, that $\gamma > 0$.  Thus, it requires a non-zero effective photon mass, such as we have in superconductivity. Note that if the condition Eq.~\eqref{stability} is relaxed, the modes at very low $k$ will still be stable since the   $ A^2$ term dominates, and the same will be true for large $k$ modes where the $B^2$ term dominates. There will however be a region of intermediate $k$ where the $\vec A \cdot\vec B $ term will give an instability. In a more complete theory this instability could be cured by higher order terms and this would open for a non-trivial magnetic structure in the ground state. We will revisit this subject in a slightly different context in the Sec.~\ref{sec:cm}.

\subsection{Connection to Optical Activity}

The optical activity of a material is usually described in terms of a frequency and momentum dependent dielectric constant and/or magnetic permeability. We shall consider the latter case and write the magnetic energy density as
\be{magenden}
{\mathcal E}_m &=& \half B^i \mu^{-1}_{ij}(\omega, \vec\nabla) B^j  \nonumber \\
&\rightarrow& \frac{1}{2\mu} B^2 + \frac {\alpha(\omega)} 2 \vec B\cdot \vec\nabla \times\vec B  \, ,
\ee
where in the second line we put $\mu_{ij}(\omega, \vec\nabla) = \mu(\delta_{ij} - \alpha(\omega) \epsilon_{ikj}\nabla^k)$ and expanded to leading order in $\vec\nabla$. The energy corresponding to the second term can be rewritten as
\be{chenergy}
E_c &=& \frac {\alpha(\omega)} 2  \int_V d^3x\, (\vec\nabla\times\vec A) \cdot  (\vec\nabla\times\vec B ) \\
&=& - \frac {\alpha(\omega)} 2  \int_V d^3x\, \vec A\cdot \nabla^2\vec B  \nonumber \\ 
&+& \frac {\alpha(\omega)} 2  \int_{\delta V} dS_i\, A_j (\partial_i B_j - \partial_j B_i ) \nonumber.
\ee
We already mentioned that the first term on the second line is not gauge invariant, but in this case it is easy to show that the surface term in the last line, as expected, restores gauge invariance, so there is no need for an additional surface theory.

Let us now assume that we have a superconductor with randomly implanted optically active impurities, that will add a term $E_c$ to the free energy functional of the superconductor. To leading order in $\alpha$, we can then just substitute the London relation $\nabla^2 \vec B = \lambda_L^{-2} \vec B$ in Eq.~\eqref{magenden}, and assuming that $\alpha (\omega)$ can be approximated by a constant $\alpha$ at low frequencies we obtain the low-energy Lagrangian in Eq.~\eqref{L_density} if we identify $\beta = \alpha(0) /\lambda_L^2$.

\subsection{Solution schema}

We are interested in solving the equation 
\begin{equation}\label{generalized_London}
\vec{\nabla} \times \vec{\nabla} \times \vB  + \beta \vec{\nabla} \times \vB + \gamma \vec{B} ~=~ 0.
\end{equation} 
Equation\,(\ref{generalized_London}) is  a generalization of the famous London equation for superconducting magnetostatics, which is the special case $\beta = 0$.  In the London equation, $\gamma$ represents the inverse square of the penetration depth.  As we now demonstrate, one can generate solutions to Eq.\,(\ref{generalized_London}) out of solutions to the London equation with a complex coefficient.  

Indeed, inserting the superposition {\it ansatz\/} 
\begin{equation}\label{ansatz_superposition}
\vB ~=~  \vB_a \, + \, \kappa \vec{\nabla} \times \vB_a 
\end{equation}
into Eq.\,(\ref{generalized_London}) leads to 
\begin{eqnarray}
{}&{}&
\left(1+ \beta \kappa \right) \vec{\nabla} \times \vec{\nabla} \times \vB_a  \, + \, \gamma \vB_a  \nonumber \\ 
{}&{}& +\kappa \vec{\nabla} \times \vec{\nabla} \times \vec{\nabla} \times \vB_a \, + \, \left(\beta + \gamma \kappa\right) \vec{\nabla} \times \vB_a \,  \nonumber \\
 {}&{}&~=~ 0,
\end{eqnarray}
and therefore when
\begin{equation}\label{ordinary_london}
\vec{\nabla} \times \vec{\nabla} \times \vB_a + \alpha \vB_a ~=~ 0,
\end{equation} 
to 
\begin{eqnarray}
{}&{}& \left[-\alpha \, \left(1 + \beta \kappa\right) \, + \, \gamma \, \right] \, \vB_a  \nonumber \\
{}&{}&\ + \left( \,  -\alpha \kappa + \beta + \gamma \kappa \, \right) \, \vec{\nabla} \times \vB_a ~=~ 0.
\end{eqnarray}
Thus, if we enforce the algebraic relations
 \begin{eqnarray}\label{algebraic_relations}
 -\alpha \, (1 + \beta \kappa) \, + \, \gamma ~&=&~ 0 \nonumber ,
 \\
 -\alpha \kappa + \beta + \gamma \kappa  ~&=&~ 0,
 \end{eqnarray}
then $\vB$ will satisfy the generalized London equation Eq.\,(\ref{generalized_London}).   

We are given $\beta, \gamma$ and seek to solve for $\alpha, \kappa$.
From Eqs.\,(\ref{algebraic_relations}) we derive a quadratic equation for $\kappa$, that is solved by 
\begin{equation}
\kappa ~=~ \frac{-\beta \, \pm \, i \sqrt {4 \gamma - \beta^2 }} {2 \gamma}.
\end{equation}
Here we see that  the realistic situation $4 \gamma - \beta^2 > 0$ brings in complex numbers.  
Having gotten $\kappa$ in terms of $\beta, \gamma$ it is straight forward to further arrive at
\begin{equation}\label{alpha_equation}
\alpha ~=~ \gamma \, + \, \frac{\beta}{2} \left( -\beta \, \mp \, i \sqrt{4\gamma - \beta^2} \,  \right).
\end{equation}

Thus, we have two complex conjugate solutions for our auxiliary inverse square penetration depth.   An immediate physical implication is that we can expect oscillations to accompany the exponential damping of fields (and currents) we usually encounter as we penetrate a superconductor.    

Ultimately we want real solutions of our field equations.  Since our auxiliary equations are linear, we can simply use the real and imaginary parts of their solutions.  Note that since the auxiliary equations are complex conjugates of one another, they both lead us to the same real fields.  

This solution scheme embodies in a precise form the concept of field-current mixing that we anticipated heuristically.  Indeed, since $\vec{\nabla} \times \vB_a$ is the London diamagnetic current associated to $\vB_a$, the solution $\vB$ defined in Eq.\,(\ref{ansatz_superposition}) is a linear combination of its field and current.  

Let us further note the curious fact that our construction in Eq.\,(\ref{ansatz_superposition}) leads to (complex-valued) fields $\vB$ that, like $\vB_a$, satisfy Eq.\,(\ref{ordinary_london}) and thus, in view of Eq.\,(\ref{generalized_London}), 
\begin{equation}
\beta \, \vec{\nabla} \times \vB ~=~ (\alpha - \gamma) \, \vB,
\end{equation}
or 
\begin{equation}
\vec{\nabla} \times \vB ~= \frac{1}{2} \left(  -\beta \, \mp \, i \sqrt{4\gamma - \beta^2} \right) \, \vB.
\end{equation}
In the critical case $4\gamma - \beta^2 = 0$, we get force-free fields.

The solutions to the generalized London equation govern the magnetostatics for parity-violating superconductors. Examples are the noncentrosymmetric superconductors \cite{PhysRevB.70.104521,AGTERBERG200313,Bauer2012NonCentrosymmetricS, PhysRevB.102.184517}, which allows for Lifshitz invariants in the Ginzburg-Landau free energy. These terms are  linear in covariant derivatives and magnetic fields, and can give rise to to helical phases and effects similar to the CME, see Ref.~\cite{Bauer2012NonCentrosymmetricS} for an extensive summary on the topic.

We now turn to the explicit solutions of the generalized London equations in various geometries with the purpose to demonstrate how pertinent effects, such as as induced currents and magnetic field profiles, are affected by the geometry of the superconductor.

\subsection{Slab geometry}
A relatively simple, yet physically significant and mathematically transparent situation to analyze is the half space or slab geometry.  Thus, we imagine our superconductor to fill the half space $x>0$ while in the remaining half-space we have a constant magnetic field 
\begin{equation}
\vec{B}^{\rm ext.} = \hat z B_0,  \ \ \ \ \ (x<0).
\end{equation}
Here we match onto the solution of the ordinary London equation Eq.\,(\ref{ordinary_london}) proportional to
\begin{equation}
\vB_a ~=~ \hat z B_0 e^{-\sqrt \alpha x} ,
\end{equation}
whose curl
\begin{equation}
\vec{\nabla} \times \vB_a ~=~ \hat y B_0 \sqrt \alpha e^{-\sqrt \alpha x} 
\end{equation}
is the diamagnetic screening current.  We will build our solution using this auxiliary form with $B_0 = 1$.

We invoke Eq.\,(\ref{alpha_equation}) to choose
\begin{equation} \label{square_root}
\sqrt \alpha ~=~ \sqrt{\gamma - \frac{\beta^2}{4}} + i \frac{\beta}{2} 
~\equiv~ p + iq,
\end{equation}
where positive square roots are understood throughout.
Note that in order to get solutions that fall off as $x \rightarrow \infty$, we must take roots with a positive real part; the remaining choice associated with the $\mp$ in $\alpha$ has no effect on our final result, and we have chosen the lower sign.

With that preparation, we can use our solution scheme to solve the generalized London equation Eq.\,(\ref{generalized_London}).  After some algebra, we arrive at 
\begin{eqnarray}
B_z ~&=&~ e^{-px} \cos qx, \nonumber \\
B_y ~&=&~ -e^{-px}  \sin qx,
\end{eqnarray} 
\begin{eqnarray}
B_z ~&=&~ - e^{-px} \sin qx ,\nonumber \\
B_y ~&=&~ - e^{-px}  \cos qx,
\end{eqnarray} 
for the real and imaginary parts.  Finally, to ensure continuity of the magnetic field at the boundary we take the linear combination of these two solutions that has $B_y(0) = 0$ and $B_z (0) = B_0$.  This gives us the magnetic field 
\begin{eqnarray}
B_z (x) ~&=&~ B_0 e^{-px} \cos qx,\nonumber \\
B_y (x) ~&=&~ - B_0 e^{-px} \sin qx,
\end{eqnarray} 
and the current
\begin{eqnarray}
&j_z(x) &= ~\frac{\partial B_y}{\partial x}  = - B_0 e^{-px} (-p \sin qx + q \cos qx),  \nonumber \\ 
&j_y(x) &= ~-\frac{\partial B_z}{\partial x} = B_0 e^{-px} (p \cos qx + q \sin qx  ),
\end{eqnarray} 
inside the superconductor.  

As anticipated, this solution displays three qualitatively new features relative to the usual London ($\beta = 0$) case.  Most profoundly, there is a current running parallel to the external field direction.  Secondly, there is an induced perpendicular magnetic field in the interior.  Thirdly, the interior fields and currents have an oscillatory character. 

We can also consider a slab, occupying the region $0 \leq x \leq a$.  We can use the same solution inside the superconductor, matched to the constant field
\begin{eqnarray}
B_z (x \geq a)~&=&~ - B_0 e^{-pa} \sin qa, \nonumber \\
B_y (x \geq a) ~&=&~ - B_0 e^{-pa}  \cos qa.
\end{eqnarray} 
This represents the result of applying $B = B_0 \hat z$ at $x\leq 0$ and a rotated (and damped) field at $x\geq a$.  Here we see a close analogy, in magnetostatics, to optical activity (accompanied by absorption).

\subsection{Intrinsic solenoid (trapped flux and model vortex)}

We can notionally insert a solenoid into our superconductor, and ask that it be generated self-consistently by screening currents within the superconductor.  This is of interest in itself, and also allows us to anticipate and model, within the relatively simple and parameter-sparse context of the (modified) London equations, properties of trapped flux and of quantized magnetic vortices. 

\begin{figure}[t!]
    \centering
    \includegraphics[width=\columnwidth]{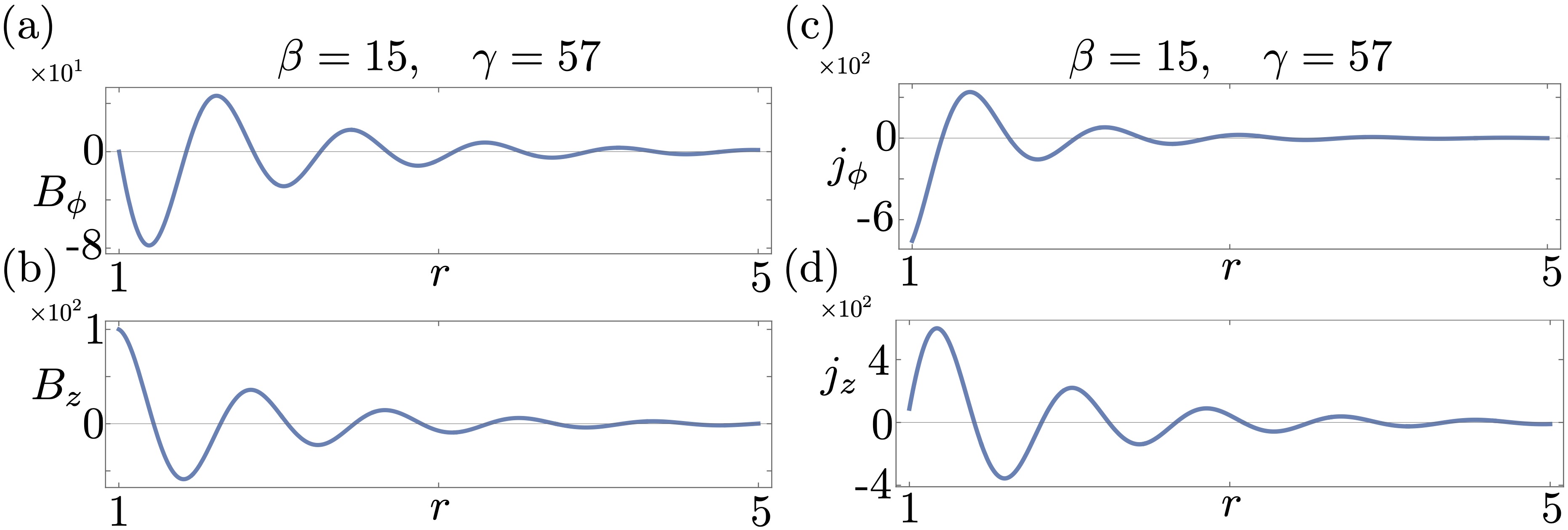}
    \caption{Azimuthal and longitudinal components of the magnetic field and the currents inside the superconductor with a solenoid for $\beta=15$, $\gamma=57$, $B_0=100$, and $R=1$, displaying a (spatially damped) longitudinal current. Notably, the applied, constant and longitudinal magnetic field in the solenoid gives rise to an azimuthal field component inside the superconductor that further oscillates in a damped fashion, just as the longitudinal component, as a function of $r$.}
    \label{fig:cylinder} 
\end{figure}

In cylindrical coordinates, our solenoid is defined by
\begin{equation}
\vB(r, z, \phi) ~=~ B_0 \hat z, \ \ \ r \leq R,
\end{equation}
and it joins on to a solution of Eq.~\eqref{generalized_London} for $r \geq R$.  The self-consistency condition is that there are no singular surface currents, which we enforce by demanding continuity of the tangential magnetic fields at $r= R$.   

\begin{figure*}[hbt!]
    \centering
    \includegraphics[width=\textwidth]{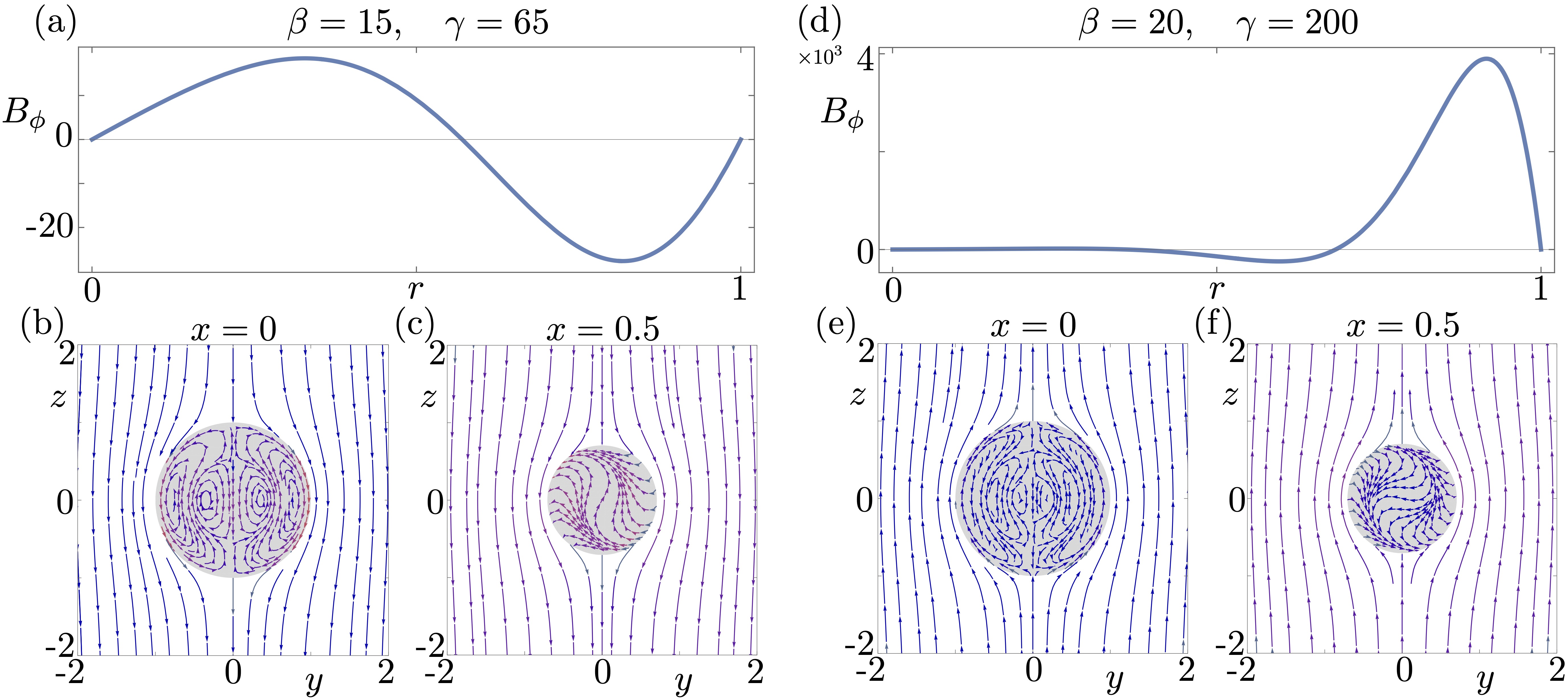}
    \caption{Expulsion of the magnetic field by a superconducting unit sphere for representative parameter values.  The magnetic field forms closed loops inside the superconducting sphere. Panels (a) and (d) display how the parity violating azimuthal component oscillates inside the sphere, for polar angle $\theta=\frac{\pi}{3}$. The field configuration gives rise to an anapole moment, as discussed in the text.}
    \label{fig:sphere} 
\end{figure*}

The auxiliary solution of the ordinary London equation brings in the Bessel function $K_1$, which dies exponentially at infinity:
\begin{equation}
B_a(r, z, \phi) ~=~ B_0 \hat z \, \frac{K_1({\sqrt \alpha} r)}{K_1({\sqrt \alpha} R)}, \ \ \  r \geq R.
\end{equation}
From the curl of this field, we infer the azimuthal diamagnetic screening current.
With this starting point, we can invoke the machinery of our solution schema to generate solutions of our generalized London equation in the exterior (superconducting) region.  Details are spelled out in Appendix~\ref{app:sol}.  

Let us mention how the qualitative novelties we observed above get manifested here:  within the superconductor we find longitudinal current flows $j_z$, azimuthal magnetic fields $B_\phi$, and oscillatory behavior (possibly damped) of all the fields and currents as functions of $r$. These general findings are displayed in Fig.~\ref{fig:cylinder} for some exemplary values of $\beta$ and $\gamma$.

It is possible to consider cylindrical shells, fields imposed from the outside, and so forth, both analytically and numerically (and, presumably, experimentally), based on the same ideas.   

\subsection{Sphere geometry}

Another accessible problem, often considered to be the paradigmatic Meissner effect,  is the superconducting sphere exposed to a constant external magnetic field.  The auxiliary reference problem here was solved and presented by London himself in his classic book \cite{London1950}. One finds, for spheres much larger than the penetration depth, the field canceled or expelled by azimuthal diamagnetic screening currents near the surface of the sphere.   Magnetic field lines with the superconductor get routed to within that penetration region, as displayed by Fig.~\ref{fig:sphere}.  In additional to the imposed field, one finds a calculable magnetic dipole arising from the circulating currents.

Since the auxiliary solution is expressed in terms of exponentials, we can use our solution schema to generate completely explicit solutions of the modified equations in terms of exponentials and trigonometric functions. The behavior of the magnetic field is illustrated in Fig.~\ref{fig:sphere} for various penetration depths, and calculational details along with the full solution are spelled out in Appendix~\ref{app:sol}.

In Fig.~\ref{fig:spherecurrents}, we see that the currents that run along the surface of the sphere and return in a (squashed) toroidal fashion give no external moment, but represent a form of what are called anapole moments in the literature. An anapole moment, or a magnetic toroidal moment, is a term in the multipole expansion of the electromagnetic field that violates both $\mathcal{P}$ and $\mathcal{T}$ symmetry. The anapole moment is given by
\begin{equation}
    T_i = \frac{1}{10}\int\left[r_i\left(\vec{r}\cdot \vec{j}\right)-2r^2J_i\right] d^3x,
\end{equation}
where $r_i$  are the Cartesian coordinates, and $\vec{j}$ the current. Using the explicit solutions of the magnetic field inside the sphere (see Appendix~\ref{app:sol} for their explicit appearance), the current is given by $\vec{j}_{\text{sphere}} = \vec{\nabla} \times \vec{B}^{\text{in}}_{\text{sphere}}$, it can be shown that $T_x$ and $T_y$ are identically zero. However, $T_z$ is finite and for the two cases illustrated in Fig.~\ref{fig:sphere} given by,
\begin{align}
    T_z\left(\beta=15,\gamma=65\right) &=-15.5804, \label{eq:anapole1}
    \\
    T_z\left(\beta=20,\gamma=200\right) &=2182.52.  \label{eq:anapole2}
\end{align}
Appendix~\ref{app:sol} contains a general expression for the $z$ component of the anapole moment.

\begin{figure}[t!]
    \centering
    \includegraphics[width=\columnwidth]{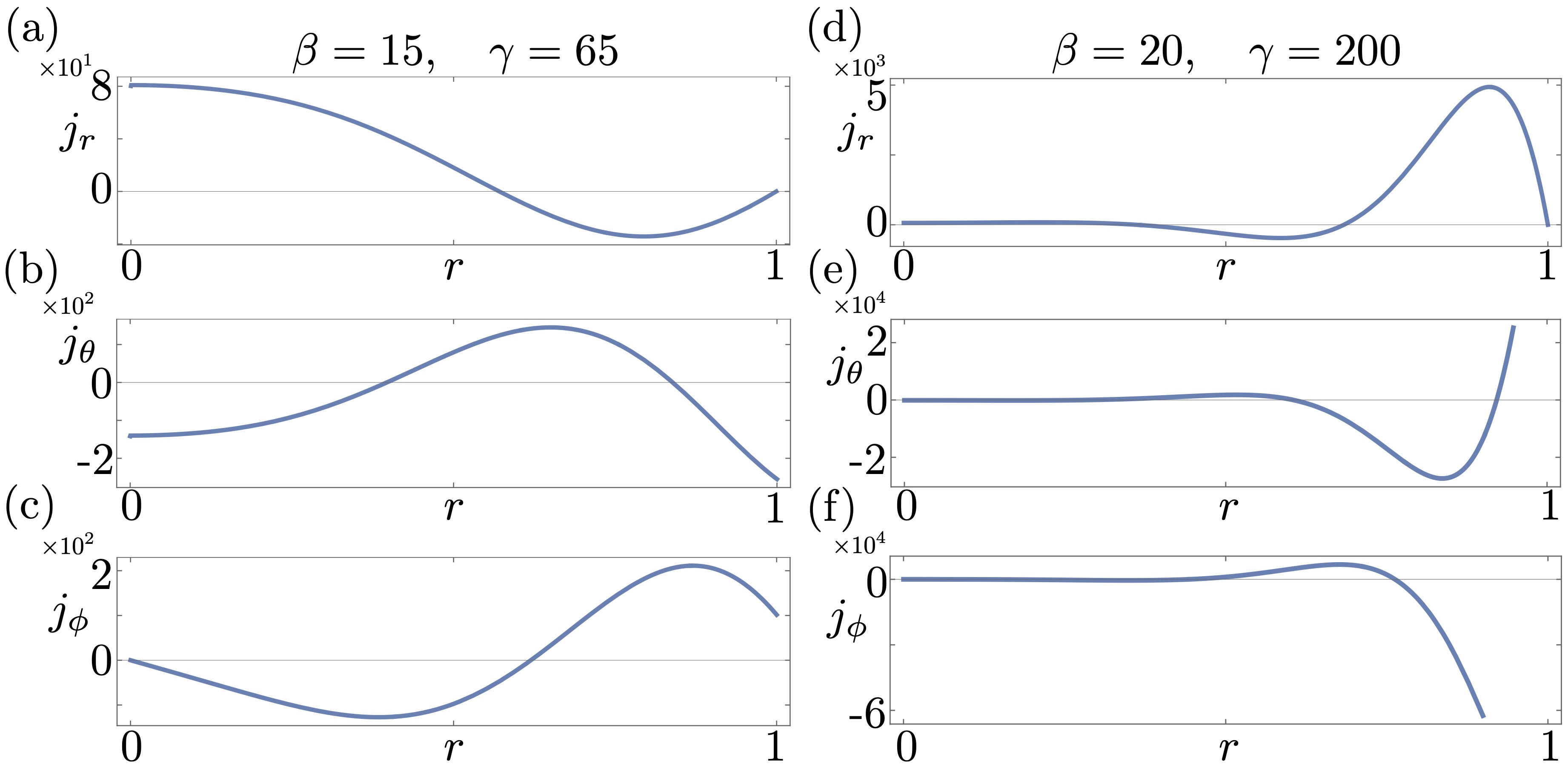}
    \caption{Currents in spherical coordinates inside a superconducting unit sphere generated by the external and constant magnetic, for polar angle $\theta=\frac{\pi}{3}$. Note that the radial component of the current vanishes at the boundary.  No current escapes from the sphere, but there is an anapole moment. }    \label{fig:spherecurrents} 
\end{figure}

\section{Actively Chiral Magnetism}\label{sec:cm}

Chiral materials generally, and not only superconductors, support a $\vB\cdot (\vec{\nabla} \times \vB)$ term in the effective Lagrangian density.  At optical frequencies, it gives optical activity.  One should also consider its effect in magneto-statics.   In that context, the most relevant terms are
\begin{equation}
-{\cal L} ~=~ \frac{1}{2\mu} B^2 + \kappa \vB\cdot \left(\vec{\nabla} \times \vB\right) + \frac{\lambda}{2}  \left(\vec{\nabla} \times \vB\right)^2.
\end{equation}
This Lagrangian density bears a close family resemblance to the effective Lagrangian we used in our analysis of chirally active superconductivity..  Indeed, the substitution $\vec{\nabla} \times \vB \rightarrow \vA$ brings it into the form we analyzed above.

A simple heuristic consideration suggests the common occurrence of this term in organic (chiral) diamagnetism.   Imagine a helical molecule segment along which diamagnetic currents can flow.   If we apply magnetic flux along the axis of the helix, the diamagnetic current that along the helix, regarded vectorially, will have a component along the applied magnetic field direction.  But the current sources $\vec{\nabla} \times \vB$, so this correlation represents a $\vB \cdot \vec{\nabla} \times \vB$.  If the magnetic field is off-axis only its component along the axis will be operative, but the same logic applies.   Helices of the same chirality will all contribute with the same sign.

We can write the energy density as 
\begin{eqnarray}
 &{}&\frac{1}{4} \left[ 
\left(\frac{1}{\mu} + \frac{\kappa}{\sqrt{\lambda \mu}}\right)\left( \vB + \sqrt{\lambda \mu} \vec{\nabla} \times \vB\right)^2\right. \nonumber \\
&{}&\ \ \ \ \  \left.+ \left(\frac{1}{\mu} - \frac{\kappa}{\sqrt{\lambda \mu}}\right)\left( \vB - \sqrt{\lambda \mu} \vec{\nabla} \times \vB\right)^2  \right],
\end{eqnarray}
from which we see that we have stability for
\begin{equation}\label{positivity}
\frac{\lambda}{\mu} \geq \kappa^2,
\end{equation}
and of course $\mu, \lambda \geq 0$.  

 Alternatively we can consider plane waves
\begin{equation}
\vA ~=~ \left(\begin{array}{c}1 \\ \pm i  \\0 \end{array}\right)e^{ikz} ,
\end{equation}
with energy density proportional to
\begin{equation}\label{quadratic_energy}
\frac{1}{\mu} \pm 2\kappa k + \lambda k^2.
\end{equation} 
Here, positivity of the energy density (for real $k$) leads again to Eq.\,(\ref{positivity}).  As long as $\lambda > 0$, we can stabilize the model by adding a $(B^2)^2$ term.  

Taking $\lambda, \kappa, k >0$, the minimum energy density taking into account only quadratic terms occurs, according to Eq.\,(\ref{quadratic_energy}), at 
\begin{equation}
k_c = \frac{\kappa}{\lambda} ,
\end{equation}
with the lower choice of sign, where it has the value 
\begin{equation}
\varepsilon ~\equiv~ \frac{1}{\mu} - \frac{\kappa^2}{\lambda}.
\end{equation}
When $\varepsilon <0$, we can lower the energy by bringing in fields of the form. 
\begin{eqnarray}\label{spiral_ferro}
B_x ~&\propto&~ 	\cos k_c z,		\nonumber \\
B_y ~&\propto&~ 	\sin k_c z,	\nonumber	\\
B_z ~&=&~ 0.
\end{eqnarray}
Note that this instability does not require $\mu < 0$, i.e., instability toward ordinary ({\it i.e.}, $k=0$) ferromagnetism, though of course it includes that possibility.  To describe a stable system, we must bring in a 
$(B^2)^2$ penalty term that limits the amplitude of the spontaneously developed structure.   Equations~(\ref{spiral_ferro}) 
represent fields that are constant within $z= {\rm const.}$ planes whose direction rotates periodically within the $x-y$ plane as $z$ varies.  In other words, we see here magnetic fields characteristic of optical activity frozen in time.

A point of interest is that because the coupling $\frac{\lambda}{2}  (\vec{\nabla} \times \vB)^2$ required to stabilize the $\kappa \vB\cdot (\vec{\nabla} \times \vB)$ optical activity term contains a larger number of derivatives than the minimal Maxwell $\frac{1}{2\mu} B^2$ term it cannot be regarded as a uniformly small perturbation, even when $\lambda$ is small.  Indeed, it changes the nature of the boundary value problem.  If we fix the gauge $\vec{\nabla} \cdot \vA =0$  and (for simplicity) set $\kappa = 0$, varying ${\cal L}$ leads to the equation
\begin{equation}\label{reference_equation}
\left[\frac{1}{\mu} + \lambda \left(\nabla^2 \right) \, \right] \, \nabla^2 \vA ~=~ 0.
\end{equation}
Any harmonic vector field will solve this equation, and very naively one might expect that for small $\lambda$ such fields provide excellent approximate solutions in general.  But in regions where $\vA$ varies rapidly the second term comes in strongly, and other solutions may be physically appropriate.  In particular, the boundary conditions one must apply at surfaces where the value of $\lambda$ changes (notably, at boundaries between our chiral magnetic materials and conventional materials, or empty space) one must enforce additional continuity of normal derivatives, beyond what is usually required for harmonic fields, and this may require substantial adjustments of candidate solutions near the boundary.  Closely related mathematical issues arise in hydrodynamics, where they have stimulated the development of boundary layer theory.

The solution schema we used in the $\vA\cdot \vB$ problem continues to work in this new context, so we can leverage known solutions of Eq.\,\eqref{reference_equation} to get solutions of the full ($\kappa \neq 0$) equations in various geometries. 

\section{Phenomenology of Axion Wind Materials}
The  axion wind term 
\begin{equation}
{\cal L}_w \ \propto \vec{\beta}_i \cdot \epsilon^{i\alpha \beta \gamma} A_\alpha F_{\beta \gamma} \ \propto \ \vec \beta \cdot (\vec A \times \vec E) - A_0 \vec \beta \cdot \vec B
\end{equation}
breaks rotation symmetry (and time-reversal symmetry), so its phenomenology is more complicated.  
Here we confine ourselves to a simple but important general observation and a calculation of its effect on wave propagation.  For simplicity, we will take $\vec \beta = \beta \hat z$.     

Whereas the term $\propto \vA\cdot \vB$ brings in spatial derivatives in all directions, and thereby involves the material as a whole, the term $\vec \beta \cdot (\vec A\times\vec E)$ does not bring in derivatives in the $\hat z$ direction, and in that sense reduces to a stack of planar terms.  (Of course, other terms in the Lagrangian will link the planes.)    Within each plane, we have, in effect, a (2+1) dimensional Chern-Simons theory.  Indeed, through the alternative formulation
$\epsilon^{3\alpha \beta \gamma} A_\alpha \partial_\beta A_\gamma$ of our term, we see that in isolation it literally represents a stack of independent (2+1)-dimensional Chern-Simons theories.  This interpretation indicates that in a bounded sample there will be massless surface modes, as in the quantum Hall effect, whose anomalies cancel the surface terms that otherwise obstruct full gauge invariance of the bulk theory.  

From the Lagrangian
\be{ake}
{\mathcal L} = \half E^2 - \half B^2 + \frac{\beta}{2}  \left[ \hat z \cdot \left(\vA \times \vec{E}\right) - A_0 B_z \right],
\ee
we derive the equations of motion
\begin{eqnarray}
\vec{\nabla} \cdot \vB ~&=&~ 0, \nonumber \\
\vec{\nabla} \times \vec{E} ~&=&~ -\frac{\partial \vB}{\partial t} ,\nonumber \\
\vec{\nabla} \cdot \vec{E} ~&=&~ \beta B_z, \nonumber \\
\vec{\nabla} \times \vB ~&=&~ \frac{\partial \vec{E}}{\partial t} - \beta \hat z \times \vec{E}.
\end{eqnarray}  
Thus we have effective charge and current densities
\begin{eqnarray}
\rho_e ~&=&~ \beta B_z, \\
\vec{j}_e ~&=&~ -\beta \hat z \times \vec{E},
\end{eqnarray}
that automatically satisfy the conservation equation. 

Having imposed the corresponding variational equation, we can set the non-dynamical field $A_0 = 0$.   

For plane waves propagating in the $\hat z$ direction, we use the {\it ansatz\/} $A = \epsilon\, e^{i(kz - \omega t)}$. We find that transverse circular polarizations lead to uncoupled dispersion relations, in the forms
\begin{align}
\vec{A} &=  \left(\begin{array}{c}1 \\ \pm i  \\0\end{array}\right)  e^{i(kx - \omega t)},
\\
0 &= \omega^2 - k^2 \mp \beta \omega .
\end{align}
Here again we find that the different circular polarizations travel at different velocities, so there is optical activity.  Unlike before, however, here we have no zone of instability.

For plane waves propagating in the $\hat x$ direction, we use the {\it ansatz\/} $A = \epsilon e^{i(kx - \omega t)}$.  One eigenmode  does not feel the new term at all:
\begin{align}
\vA &=  \left(\begin{array}{c} 0 \\ 0 \\ 1 \end{array}\right)  e^{i(kx - \omega t)},      
\\
0 &= \omega^2 - k^2.
\end{align}
The other eigenmode is more unusual.  It is 
\begin{equation}
\vA ~=~       \left(\begin{array}{c} i\frac{\beta}{\omega} \\ 1  \\0\end{array}\right)  e^{i(kx - \omega t)},        
\end{equation}
with the dispersion relation
\begin{equation}
0 ~=~ \omega^2 - k^2 - \beta^2.
\end{equation}
This dispersion relation is characteristic of a massive excitation.  The polarization, which is never transverse, becomes increasingly longitudinal at low frequencies.

It is straightforward, though lengthy, to calculate the general case  $\vec k = k(\sin \theta \hat x + \cos \theta \hat z)$. Here we only record the dispersion relation:
\begin{equation}
\omega^2 ~=~ k^2 \, + \frac{\beta^2}{2} \pm \beta \sqrt{k^2 \cos^2 \theta + \frac{\beta^2}{4}}.
\end{equation}

\begin{figure}[t]
\includegraphics[width=0.65\columnwidth]{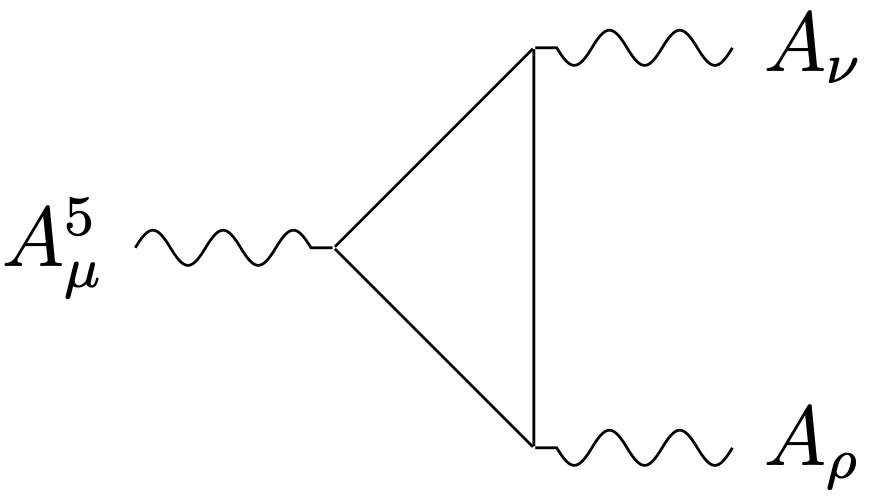}
\caption{Given an effective axial vector potential, two-photon response includes the classic VVA triangle anomaly graph.}
 \label{fig:triangle}
\end{figure}

\section{Microscopic Models Based on Semimetals} \label{sect:micro}

Previously, we discussed, in general terms several situations where we can expect emergent Chern-Simons terms to arise.  In this section, we will describe in detail a specific construction, inspired by the appearance of anomalies in quantum field theories, that gives rise to them.  

The $\mathcal{P}$ or $\mathcal{T}$-breaking terms we seek are quadratic in gauge potentials and involve cross products.  A natural way for those to arise, in Feynman graphs, is through the structure $\Tr[ (\vec a \cdot \vec\sigma)  (\vec b \cdot \vec\sigma)  (\vec c \cdot \vec\sigma)]$ when integrating over fermions in vacuum polarization loops.  This is similar to the structure that gives rise to chiral anomalies in relativistic theories in even dimensions. Inspired by these thoughts, we are led to consider Dirac and Weyl materials.   

\subsection{Axion wind terms in Weyl semimetals}

In WSMs the different chiralities can be split in momentum space by imposing stress. The vector separating the nodes is axial, and thus provides a candidate for the axial vector $\vec k$. We now show, by an explicit calculation of the vacuum current, that such a splitting does lead to a $\sim \vec k \cdot (\vec A\times\vec E)$ term in $S_{\text{eff}}$  \cite{HOSUR2013857}.

Consider an inversion symmetric Weyl semimetal with Hamiltonian
\begin{align}
\mathcal{H} &= \sum_{a=\pm}\sum_{\vq}c_a^{\dagger}(\vq_a) h_a(\vq_a)c_a  \nonumber
\\
&= \sum_{\vq}\tilde{\Psi}^{\dagger}(\vq) \begin{pmatrix} h_-(\vq_-) & 0 \\0& h_+(\vq_+)\end{pmatrix}\tilde{\Psi}(\vq),
\end{align}
where
\begin{align}
h_{\pm}(\vq^{\pm}) &= q^{\pm}_x \sigma^x+q^{\pm}_y \sigma^y+q^{\pm}_z \sigma^z - b_0^{\pm} \sigma^0,
\\
 \vq_{\pm} &=\left[k_x\pm b_x,k_y\pm b_y,\mp\left(k_z\pm b_z\right)\right],
\end{align}
and the wave functions are defined as
\begin{equation}
\tilde{\Psi}(\vq) = \begin{pmatrix} c_-(\vq)\\c_+(\vq)\end{pmatrix}.
\end{equation}
The Weyl semimetal has a pair of nodes located at $\vec{k}=\pm \vec{b}$ at energies $b_0^{\pm}$. Here, we have set the product of Planck's constant and the Fermi energy, $\hbar v_F=1$. In the following, to simplify notation the sum over momenta will be left implicit. 

The Hamiltonian can be re-written as
\begin{align}
\mathcal{H} &= \tilde{\Psi}^{\dagger} \begin{pmatrix} \left( \vk + \vb \right) \cdot \vsigma +b_0^- & 0 \\ 0 & \sigma^z \left[ -\left(\vk-\vb\right)\cdot \vsigma +b_0^+\right] \sigma^z\end{pmatrix} \tilde{\Psi} \nonumber 
\\
&= \Psi^{\dagger} \begin{pmatrix} \left( \vk + \vb \right) \cdot \vsigma +b_0^- & 0 \\ 0 & \  -\left(\vk-\vb\right)\cdot \vsigma +b_0^+\end{pmatrix}\Psi,
\end{align}
with $\Psi(\vq) = \left(c_-(\vq),\sigma^z c_+(\vq)\right)^{\text{T}}$. Using the chiral representation of the Weyl matrices,
\begin{equation}
\gamma^0 = \begin{pmatrix} 0&\sigma^0\\ \sigma^0& 0 \end{pmatrix}, \quad \vec{\gamma} = \begin{pmatrix} 0& \vsigma \\ -\vsigma & 0 \end{pmatrix}, \quad \gamma^5 = \begin{pmatrix} -\sigma^0 &0 \\ 0 & \sigma^0 \end{pmatrix},
\end{equation}
the Hamiltonian becomes
\begin{align}
\mathcal{H} &= \bar{\Psi} \left[-\vgamma\cdot \vk + \vgamma\cdot \vb \gamma^5 + \gamma^0 \left( P_-b_0^-+P_+b_0^+\right)\right] \Psi,
\end{align}
with $P_{\pm} = \frac{1}{2}\left({\bf 1}\pm \gamma^5\right)$ and $\bar{\Psi}= \Psi^{\dagger}\gamma^0$. Heisenberg equations of motion then yield
\begin{align}
\left[ \slashed{\partial}-\vgamma\cdot \vb \gamma^5 - \gamma^0 \left(P_-b_0^-+P_+b_0^+\right)\right]\Psi&=0 \, ,
\end{align}
from which we can extract the Lagrangian:
\begin{equation}
\mathcal{L} = \bar{\Psi}\left[ i \slashed{\partial}-\vgamma\cdot\vb \gamma^5 - \gamma^0\left(P_-b_0^-+P_+b_0^+\right)\right] \, .
\end{equation}
Coupling to the electromagnetic field via minimal coupling gives 
\begin{equation}
\mathcal{L} = \bar{\Psi} \left[ i \slashed{\partial} + \slashed{A} -\vgamma\cdot \vb \gamma^5 -\gamma^0\left(P_-b_0^-+P_+b_0^+\right)\right] \Psi \, .
\end{equation}
By shifting the zeroth component of the gauge field $A$ such that $A_0\to A_0-\frac{1}{2}\left(b_0^++b_0^-\right)$, and defining $b_0 = \frac{1}{2}\left(b_0^--b_0^+\right)$, it can be written as
\begin{equation}
\mathcal{L} = \bar{\Psi}\left( i\slashed{\partial} + \slashed{A}+\slashed{b}\gamma^5\right) \Psi.
\end{equation}

Thus, the shifts in momentum and energy of the Weyl nodes can be recast into an effective axial gauge field (of a very special form) in the Lagrangian. 
A constant $b_\mu$ can be written $b_\mu = \partial_\mu \xi$, with $\xi = b_\mu x^\mu$. Naively, such a $b_\mu$ can be eliminated by the chiral rotation $\psi \rightarrow e^{i\xi\gamma^5}\psi $.  However, a nonzero contribution to the effective action arises from the triangle anomaly, Fig.~\ref{fig:triangle}.  After a partial integration, it becomes (see, e.g., Ref.~\cite{nair06})
\begin{equation}
 S_{\text{top}} = \frac{1}{8\pi^2} \int d^4x \epsilon^{\mu\nu\rho\sigma} \Tr \left(b_{\mu}A_{\nu}\partial_{\rho}A_{\sigma}\right).
\end{equation}
Here, $\epsilon^{\mu\nu\rho\sigma}$ is the Levi-Civita symbol with convention $\epsilon^{0123}=1$. We will now break down the individual components of this variation and write them out explicitly to understand their physical significance:

\begin{widetext}
\begin{align}
 S_{\text{top}} &= \frac{1}{8\pi^2} \int d^4x\left( \epsilon^{0ijk}b_0A_i\partial_jA_k + \epsilon^{i0jk} b_iA_0\partial_jA_k+\epsilon^{ij0k}b_iA_j\partial_0A_k+\epsilon^{ijk0}b_iA_j\partial_kA_0\right) \nonumber
\\
&= \frac{1}{8\pi^2}\int d^4x \left(  b_0 \vec{A}\cdot \vec{B} -A_0 \vb\cdot \vec{B} + \vb \cdot \vec{A} \times \vec{E} \right). \label{eq:effactWSM}
\end{align}
\end{widetext}
Here, we again used that $b_{\mu}$ are constants, in order to perform integration by parts.

At first glance it seems that an $\vec{A}\cdot \vec{B}$ term appears already here, and its presence has been debated in recent years \cite{Landsteiner,HOSUR2013857,PhysRevB.102.241401,10.21468/SciPostPhys.10.5.102,PhysRevLett.111.027201,PhysRevB.89.035142,PhysRevB.88.125105,PhysRevB.91.115203}. Here, however, we have calculated the current in the vacuum, meaning that we have assumed that each of the Weyl cones are filled up exactly to the nodes, despite them being at different energies. Having finite density amounts to filling particles at the nodes to levels set by the chemical potentials $\mu_+$ and $\mu_-$, and the total current becomes  $\vec{j}=\frac{\mu_5-b_0}{2\pi^2}\vec{B}$, where $\mu_5 = \mu_+-\mu_-$, and the energy shift $b_0$ plays the role of an $A_0^5$-term \cite{Landsteiner}.
In realistic systems, this is not a stable situation, since there are scattering processes that  transfer electrons between the cones, so the (generalized) grand  canonical equilibrium amounts to having both Weyl cones  filled up to the same Fermi energy rather than to the separate ones. This amounts to $\mu_5 = A_0^5$ and thus no CME, so we are left with
\begin{equation}
 S_{\text{top}} = \frac{1}{8\pi^2} \int d^4x \left(\vb \cdot \vec{A} \times \vec{E}-A_0\vb\cdot\vec{B}\right),
\end{equation}
where $\vb$, as promised,  is the (constant) axial vector that gives rise to the $\vb\cdot \vec{A}\times \vec{E}$ term in $S_{\text{eff}}$. Note that there is an additional term $\sim \vb\cdot \vec{B}$, whose strength is determined by $A_0$. 

\subsection{$ k_0 \vec A\cdot\vec B$ term from a flux biased Weyl supercondutor}
It is more difficult to generate an $S_{\text{top}}[\vec E, \vec B]$ with an $\vec{A}\cdot\vec{B}$ term by this mechanism. Formally, in the static limit, such a term corresponds to an imbalance between Weyl nodes of positive and negative chirality, which is disallowed in a system with bounded energy bands according to fermion doubling theorems. Here, we further want to clarify an important difference to the work by Zhong {\it et. al.} in Ref.~\cite{PhysRevLett.116.077201}, where the current associated to the $\vec{A}\cdot\vec{B}$ term remains finite. This is due to the presence of an electromagnetic wave, resulting in a non-trivial electric field and this a driving in the system. Here, we consider the static case where we assume a constant magnetic field and no electric fields.

Imbalances between chiral nodes are, however, known to arise in several contexts including Floquet systems \cite{PhysRevLett.121.196401,PhysRevLett.123.066403,PhysRevResearch.2.033045}, and situations where effects that only gap out nodes of one particular chirality are present. Examples of the latter include chirality locking charge density waves \cite{NatComm.12.2914} and certain Weyl superconductors \cite{PhysRevLett.118.207701,PACHOLSKI2020168103,PhysRevB.104.035109}. To illustrate that terms on the form $\vec{A}\cdot \vec{B}$ indeed do arise in physically realizable systems, we will show it for a single Weyl node Hamiltonian originating from a flux-biased Weyl superconductor. 

For the sake of completeness, we first outline how the considered system is set up, referring the reader to the very insightful original work in Ref.~\cite{PhysRevLett.118.207701} for further details. The parent Hamiltonian is taken as,
\begin{align} \label{parent}
\mathcal{H} &= \sum_{\vk} \Psi^{\dagger}_{\vk}H(\vk)\Psi_{\vk}, \quad \Psi_{\vk} = \left(\psi_{\vk}, \sigma^y \psi^{\dagger}_{-\vk}\right),
\\
H(\vk) &= \begin{pmatrix} H_0(\vk - e\vec{A}) & \Delta_0 \\ \Delta_0^* & -\sigma^y H_0^*(-\vk - e \vec{A}) \sigma^y \end{pmatrix},
\\
H_0(\vk) &= \sum_i \tau^z \sigma^i \sin k_i + \tau^0\left(\beta\sigma^z-\mu\sigma^0\right) + m_{\vk}\tau^x\sigma^0,
\\
m_{\vk} &= m_0 + \sum_i(1-\cos k_i).
\end{align}
$\tau_i$ and $\sigma_i$, $i=x,y,z$, are orbital and spin Pauli matrices, respectively, $\beta$ a magnetization, $\mu$ a chemical potential, $\vec{A}$ the electromagnetic vector potential, and $\Delta_0$ the BCS-pairing potential. A system with this Hamiltonian can be obtained by stacking alternate layers of topological insulators and conventional BCS superconductors, which introduces a coupling between the Weyl nodes centered at $(0,0,\pm \sqrt{\beta^2-m_0^2})$ of $H_0$ and their corresponding particle-hole conjugates. The number of ungapped particle-hole conjugate Weyl cones determine the topological phase of the Weyl superconductor, phases that can be accessed in an externally controllable way, as explained in Ref.~\cite{PhysRevLett.118.207701}. This is done by coupling a {\em flux-bias circuit} to the material slab. This will alter the Hamiltonian, as the flux bias will be taken into account as a constant shift in the vector potential $\vec{A}$. In this particular setup of, the flux bias gives a contribution $\Lambda/e$ to $A_z$. As a result, the Weyl nodes appear at $(0,0,b_{\pm})$ and $(0,0,-b_{\pm})$, with
\begin{equation}
b_{\pm}^2 = \left(\sqrt{\beta^2-m_0^2}\pm \Lambda\right)^2-\Delta_0^2,
\end{equation}
meaning that when 
\begin{equation}
\left|\sqrt{\beta^2-m_0^2}-\Lambda\right| < \Delta_0 < \sqrt{\beta^2-m_0^2}+\Lambda,
\end{equation}
one of the two pairs of particle-hole conjugate Weyl nodes are gapped out, leaving only the nodes with positive chirality.

In the regime where Weyl nodes of only one  chirality are gapped out, we are left with two nodes of the same chirality. The contribution from the current from these respective nodes will then add up instead of canceling one another (as for nodes of opposite chirality). 
The two nodes are described by essentially the same Hamiltonian, 
\begin{equation} \label{eq:projham}
\tilde{\mathcal H}_\a = \sum_{\vk}\tilde\psi^\dagger_{\vk} \left[ \sum_i \nu_i(\delta k_i - Q_i A_i )\sigma^i - Q_0 \mu \sigma^0 \right] \tilde\psi_{\vk},
\end{equation}
where, $\vk = (0,0,b_\a) + \delta\vk$, $\vec\nu = (1,1,-\k)$, $Q_0 = \k$, ${\vec Q} = e(\k,\k,1/\k)$, and 
\be{kappa}
\k \approx \sqrt{1-\frac{\Delta_0^2}{(\beta+\Lambda)^2}}.
\ee
Using standard procedure, this can be recast as a Lagrangian, 
\be{chirlag}
\mathcal{ L}_\a = \bar\psi_\a (i\tilde{\slashed\partial} + \tilde{\slashed A}_\a  )\psi_\a,
\ee
where $\tilde{\slashed\partial} = \gamma^0\partial_0 - \nu_i \gamma^i\partial_i$ and the  left-handed chiral gauge field is $\tilde A_\a = (A_0-Q_0\mu , \, \nu_iQ_i A_i - \nu_i b_{\a;i})$. 

We now set $A_0 = 0$, and take $\vec b_\a$ to be constant, but allow for an $\vec x$-dependent chemical potential $\mu(\vec x)$.
The left-handed chiral anomaly is $\tilde\partial_\mu J^\mu = - \frac {e^2} {32\pi^2} \epsilon^{\mu\nu\sigma\lambda}    \tilde F_{\mu\nu} \tilde F_{\sigma\l} $, where $\tilde F$ is the field strength related to $\tilde A$. From this, we can extract the topological part of the chiral current:
\be{topcurr}
J^i =  - \frac {e^2} {16\pi^2} \epsilon^{i0jk}\tilde A_0\tilde\partial_i \tilde A_k  \, .
\ee
where $\tilde A_0 = -Q_0\mu$.
This expression is well-defined and finite also for constant $\mu$, but a direct calculation of $J^i$ will give a logarithmically divergent result, as shown in  Appendix~\ref{app:derivation}. This is no contradiction with \eqref{topcurr}, since for constant $\mu$ $\partial_i J^i =0$, there can be an extra contribution that is not determined by the anomaly. We believe that the limiting procedure $\partial_i \mu \rightarrow 0$, which parallels the derivation in Ref.~\cite{PhysRevLett.118.207701}, gives the correct result.  It is moreover consistent with a physically motivated  subtraction procedure, which is explained in Appendix \ref{app:derivation}.

We now express the components of the current in terms of the 
original fields $\vec A$ and $\vec B$. Recalling that we have two nodes, and that $J\equiv J_{em} = 2 J_L$, we get (for details, see Appendix \ref{app:derivation})
\begin{align} 
J^x &=  \frac {\k e^2\mu} {4\pi^2}  \left[ B^x + \left(1-\k^2 \right) \partial_z A_y \right],  \nonumber
\\
J^y &=  \frac {\k e^2\mu} {4\pi^2}  \left[B^y -  \left(1-\k^2 \right) \partial_z A_x \right], \label{eq:currents}
\\
J^z &=  \frac {\k e^2\mu} {4\pi^2} B^z \, . \nonumber
\end{align}
Integrating Eqs.~\eqref{eq:currents} we get the topological 
action,
\begin{equation} \label{stop}
S_{\text{top}}[A] =   - \frac {\k e^2\mu} {8\pi^2}\int d^3x\,  \left[ \vec A\cdot \vec B + 2  \left(1-\k^2 \right) A_x \partial_z A_y\right] ,
\end{equation}
which is the central result of this section.

The electromagnetic response in this regime can be expected to support a CME, i.e., a current in the direction on the externally applied magnetic field, which is often read off directly from the anomaly equation.  In our approach the dynamics is determined by considering the full action, as in Sect. \ref{sect:abphen}, where we do find qualitative effects of that kind. In prospective experiments it would be natural to take the field perpendicular to the stacked planes, and to require that the slabs should  be thinner than the penetration depth, so averaging over layers is justified. 

The term  $\sim A_x \partial_z A_y$ is not gauge invariant, and contrary to  $\sim \vec A\cdot\vec B$, not even in the bulk. Since we started from a gauge invariant theory, this means that in a full calculation the gauge invariance would be restored by extra terms that would effectively amount to the  substitution $2e \vec A \rightarrow 2e \vec A + \vec\nabla \phi$, where $\phi$ is the  phase of the superconducting order parameter. Using Eq.~\eqref{kappa}, the amplitude of the term  $\sim A_x \partial_z A_y$ is $\sim \frac{\Delta_0^2}{(\beta+\Lambda)}$ and could be made small in certain parameter ranges.

In  Appendix~\ref{app:derivation}, we shall give an alternative derivation of Eq.~\eqref{stop}, which is more in line with the derivation in this paper and does not  rely on the chiral anomaly. 

We note that the expression for $J^z$ is the same as derived in Ref.~\cite{PhysRevLett.118.207701} and that in Ref.~\cite{PhysRevLett.118.207701}, it is shown that this current component is  accompanied by a surface counterflow sourced by Fermi arcs. This is to ensure that the system does not have a thermal current in the ground state, as required by thermodynamics. Although we have not done an explicit calculation, we believe that the same is true for the other components in Eqs.~\eqref{eq:currents}.

We should also note that there is a hidden assumption in the preceding derivation, in that we neglected the possibility of adding a Wess-Zumino counter term, but took the perturbative result at face value. Usually, this ambiguity is fixed by requiring gauge invariance, but in our superconducting context it is less clear. 
Still, it is reassuring that the simple argument based on the anomaly gives the same result as the direct calculation in Appendix~\ref{app:derivation} if we there make a physically motivated subtraction inspired by the treatment in Ref.~\cite{PhysRevLett.118.207701}. Importantly, it was shown there that the expression for $J^z$ (which was the only one considered in that paper) agrees numerically with a direct calculation in the parent eight-band theory,  Eq.~\eqref{parent}. Although we believe that the connection to the anomaly cannot be a coincidence, we presently lack a sound theoretical argument excluding any Wess-Zumino term.

\section{Summary and Outlook}

We have motivated the consideration of emergent Chern-Simons interactions in 3+1 dimensions, displayed some of their striking phenomenological consequences, and indicated how they might be realized in plausible material systems.   

These interactions arise in two forms: $\vA\cdot \vB$ and $\hat n \cdot \vA \times \vec{E} - A_0 \hat n \cdot \vB$.    
Highlights for the first type include a precise form of current-field mixing, non-dissipative complex penetration depths, and anapole moments.    This type generically arises in $s$-wave superconductors that break parity symmetry.   It is closely related, at a mathematical level, to optical activity.  We suggest that it can be achieved in chirally purified organic superconductors or, more generally, by chiral doping or through parity-violating crystalline structures.  We also calculated its appearance in a microscopic model based on superconducting Weyl semimetals, where it arises through a mechanism closely related to the chiral anomaly of quantum field theory.  

Highlights for the second type include massless boundary excitations and unusual bulk effects in electromagnetic wave propagation.  This type appears to be comparatively easy to achieve in the Weyl semimetal context.  

We also extended the optical activity analogy in a slightly different direction---formally, towards higher rather than lower orders of gradient---to define actively chiral magnets.  These do not bring in Chern-Simons terms, but physical intuition and mathematical techniques carry over.  This extension frees us of the constraints of superconducitivity (notably, cryogenic temperatures and magnetic screening) and opens up many possibilities for realization in organic magnetism and metamaterials, as well as naturally occurring materials.   

The next, crucial development for this work will be to bring its mathematical paradise down to earth in concrete material realizations.

\vskip 3mm
{\bf Acknowledgement}: M.S. and T.H.H. thanks J. Hannukainen and J.H. Bardarson for insightful discussions. MS acknowledges fruitful discussions with E.J. Bergholtz at an early stage of this project.  F.W. is supported by the U.S. Department of Energy under Grant Contract  No. DE-SC0012567, by the European 
Research Council under Grant No. 742104, and by the Swedish Research Council under Contract No. 335-2014-7424.

\clearpage

\begin{widetext}
\appendix
\section{ALTERNATIVE DERIVATION OF EQS.~\eqref{eq:currents} AND \eqref{stop}  } \label{app:derivation}

\begin{figure}
\includegraphics[width=\linewidth]{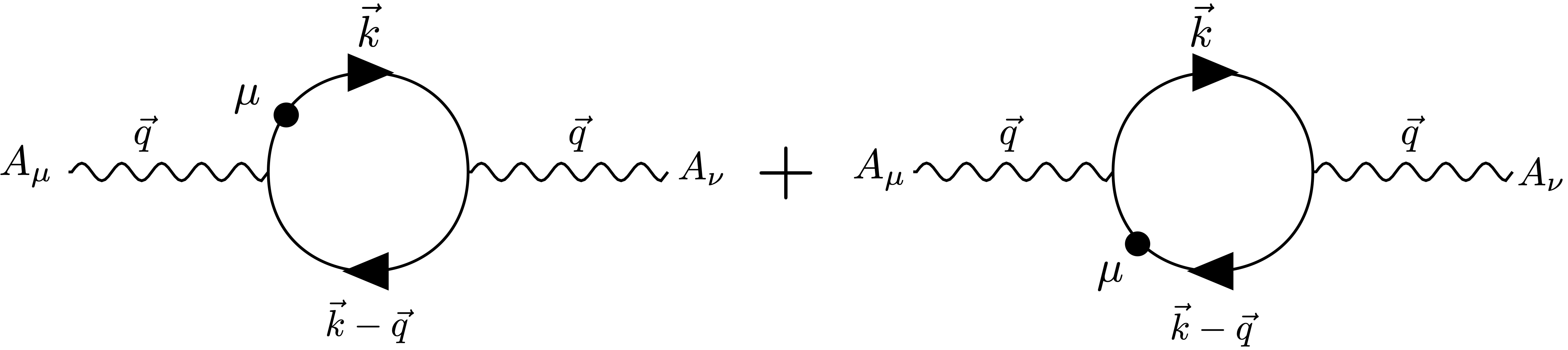}
\caption{Feynman diagrams for the polarization tensor to leading order in $\mu$.}
 \label{fig:bubble}
\end{figure}
In this appendix we recalculate Eq.~\eqref{stop} using straight-forward diagrammatic perturbation theory.
To get the pertinent static electromagnetic response function to quadratic order, we evaluate the Feynman diagram in Fig.~\ref{fig:bubble} to linear order in $\mu$ and $\vq$. The relevant integrand is,
\be{integ}
\Tr \left[ \sigma^i \frac 1 {\slashed k}  \frac 1 {\slashed k} \sigma^j  \frac 1 {\slashed k - \slashed q} +
 \sigma^i \frac 1 {\slashed k}   \sigma^j  \frac 1 {\slashed k - \slashed q} \frac 1 {\slashed k - \slashed q} \right] =
 \frac 1 {k^2} \Tr\left[ \sigma^i\sigma^j(\slashed k - \slashed q) + \sigma^i \slashed k\, \sigma^j\right] \frac 1 {(k-q)^2}
 = -\frac{ 2i} {k^4}\epsilon^{ijk} q_k + O(q),
 \ee
where we used the four-vector notation $k_\mu = (\omega, \vec k)$ and the cyclic property of the trace. Restoring $\mu$,    the energy and momentum integrals, and the minus sign due to the fermion loop, we get the polarization tensor:
\begin{equation} \label{poltens}
\Pi^{ij} = 2i\mu q_k \epsilon^{ijk}\int\frac{d^3k}{(2\pi)^3}\int\frac{d\omega}{2\pi} \frac{1}{(\omega^2+k^2)^2}.
\end{equation}  

The appearance of $\Pi^{ij}$ makes it clear that the contribution from two nodes of positive chirality symmetrically shifted from the origin of momentum space will indeed add up instead of cancel. This can be seen by shifting $k\to k\pm b$, and make an expansion for small $b$. 

The integral in Eq.~\eqref{poltens} is logarithmically divergent both in the infrared and the ultraviolet. The infrared divergence is clearly a result of expanding to order $\mu$, and is regulated if the full $\mu$-dependence is kept. However, to follow as closely as possible to Ref.~\cite{PhysRevLett.118.207701}, we shall instead regulate the infrared by a finite temperature $T=1/k_B \beta$. The ultraviolet divergence is a consequence of that, naively, there is a contribution to the current from the whole Dirac sea, as will be discussed below.
 
Using the standard Euclidean formulation of finite temperature  QFT, the polarization tensor Eq.~\eqref{poltens} at temperature $T$, becomes,
\begin{equation} \label{finitet}
\Pi^{ij} = 2i\mu q_k \epsilon^{ijk}\int\frac{d^3k}{(2\pi)^3}\frac{1}{\beta}\sum_n \frac{1}{(\omega^2_n+k^2)^2},
\end{equation}
where $T=\frac{1}{k_{\text{B}}\beta}$, $k_{\text{B}}$ is the Boltzmann constant, and $\omega_n = \frac{2\pi}{\beta}\left(n+\frac{1}{2}\right)$ are the fermionic Matsubara poles. Rewriting the sum in Eq.~\eqref{finitet} as
\begin{equation}
\frac{1}{\beta}\sum_n\frac{1}{(\omega_n^2+k^2)^2} = \frac{1}{\beta}\left(-\frac{1}{2k}\right)\frac{\partial}{\partial k}\sum_n\frac{1}{\omega^2_n+k^2}= \frac{1}{\beta}\left(-\frac{1}{2k}\right) \frac{\partial}{\partial k}\frac{\beta^2}{(2\pi)^2}\sum_n \frac{1}{(n+\frac{1}{2})^2+\left(\frac{\beta k}{2\pi}\right)^2}.
\end{equation}
and using  $\sum_{n=-\infty}^{\infty} \frac{1}{(n+\frac{1}{2})^2+(Ax)^2} = \frac{\pi}{A x} \tanh(A\pi x)$, the finite $T$ polarization tensor becomes, 
\begin{equation}
\frac{1}{\beta}\sum_n\frac{1}{(\omega_n^2+k^2)^2} =\left(-\frac{1}{2k}\right)\frac{\partial}{\partial k} \frac{1}{2k}\tanh\left(\frac{\beta k}{2}\right) = \frac{\sech^2\left(\frac{\beta k}{2}\right)\left[\sinh\left(\beta k\right)-\beta k\right]}{8 k^3}.
\end{equation}
Since the integrand is isotropic, we use spherical coordinates and $\int \frac{d^3k}{(2\pi)^3}=\int \frac{dk}{2\pi^2} k^2$, and after inserting the Jacobian factor relating $\tilde A$ to $A$ in the integral measure, we get
\begin{equation}\label{eq:finintpol}
\Pi^{ij} = 2iQ_0\mu q_k \epsilon^{ijk} \frac{1}{8\pi^2}\int_0^{\infty}\frac{dk}{|\nu_x\nu_y\nu_z|}\left[ \frac{2\tanh\left(\frac{\beta k}{2}\right)}{k}-\frac{\beta}{2}\sech^2\left(\frac{\beta k}{2}\right)\right].
\end{equation}
The first term Eq.~\eqref{eq:finintpol} is divergent, while the second is convergent. Subtracting the divergent piece, we arrive at the final result for the polarization tensor,
\begin{equation}
\Pi^{ij} = \frac{2iQ_0\mu \nu_k q_k \epsilon^{ijk}}{|\nu_x\nu_y\nu_z|}\left(-\frac{1}{8\pi^2}+ \text{Div}\right).
\end{equation}
This subtraction will be discussed below.
The finite part of the effective action, after substituting $q_k\to -i\partial_k$, becomes
\begin{equation}
S_{eff}[A] = \frac{1}{2}\nu_iQ_i A_i \Pi^{ij} \nu_jQ_j A_j =-\frac{Q_0\mu \nu_iQ_i\nu_jQ_j\nu_k}{8\pi^2|\nu_x\nu_y\nu_z|}A_i\partial_kA_j\epsilon^{ijk}=-\frac{1}{8\pi^2}\sgn{(\nu_x\nu_y\nu_z)}Q_0\mu Q_iQ_jA_i\partial_k A_j\epsilon^{ijk} \, ,
\end{equation}
and, finally, the current
\begin{align}
J^{l} &= \frac{\delta S_{eff}[A]}{\delta A_{l}} \nonumber
\\
&= -\sgn{(\nu_x\nu_y\nu_z)}\frac{\mu  Q_iQ_j}{8\pi^2}\partial_k A_j \epsilon^{ijk} \frac{\delta A_i}{\delta A_{l} }\nonumber
\\
&=-\sgn{(\nu_x\nu_y\nu_z)}\frac{\mu Q_{l}Q_j}{8\pi^2}\partial_k A_j \epsilon^{ljk},
\end{align}
where the index $l$ in the right hand side is not summed over.

Inserting the system parameters from the Hamiltonian in Eq.~\eqref{eq:projham} gives
\begin{align}
S_{\text{top}}[A] &= -\frac{\mu \kappa}{8\pi^2}Q_iQ_kA_i\partial_j A_k \epsilon^{ijk} \nonumber
\\
&= -\frac{\mu \kappa}{8\pi^2}\left[Q_xA_x \left(Q_z\partial_yA_z-Q_y\partial_zA_y\right)+Q_yA_y\left(Q_x\partial_zA_x-Q_z\partial_xA_z\right)+Q_zA_z\left(Q_y\partial_xA_y-Q_z\partial_yA_x\right)\right] \nonumber
\\
&=  -\frac{\mu e^2\kappa}{8\pi^2}\left[A_x\left(\partial_yA_z-\kappa^2\partial_zA_y\right)+A_y\left(\kappa^2\partial_zA_x-\partial_xA_z\right) + A_z\left(\partial_xA_y-\partial_yA_x\right)\right] \nonumber
\\
&= -\frac{\mu e^2\kappa}{8\pi^2}\left\{ A_x \left[\partial_yA_z-\partial_zA_y+\left(1-\kappa^2\right)\partial_zA_y\right]+A_y\left[\partial_zA_x-\partial_xA_z-\left(1-\kappa^2\right)\partial_zA_x\right]+A_z\left(\partial_xA_y-\partial_yA_x\right)\right\} \nonumber
\\
&= -\frac{\mu e^2\kappa}{8\pi^2}\left[-\vec{A}\cdot\vec{B}+\left(1-\kappa^2\right)\left(A_y\partial_zA_x-A_x\partial_zA_y\right)\right] \nonumber
\\
&=  \frac{\mu e^2\kappa}{8\pi^2}\left[\vec{A}\cdot\vec{B}+\left(1-\kappa^2\right)\left(A_y\partial_zA_x-A_x\partial_zA_y\right)\right] \, ,
\end{align}
and finally restoring $\hbar$, using the notation $e^* = \kappa e$,  the  current becomes
\begin{equation}
J^l = \frac{\delta S_{\text{top}}[A]}{\delta A_l} = \frac{\mu ee^*}{h^2}\left[ B^l + \left(1-\kappa^2\right)\left(\delta^l_y\partial_zA_x -\delta^l_x\partial_zA_y\right)\right]  .
\end{equation}
Using the gauge where $\vec{A} = \left(0,Bx,\Lambda/e\right)$, the only surviving component of the current reads,
\begin{equation}
J^z  = \frac{ee^*\mu}{h^2}B_z.
\end{equation}
which agrees with the result of Ref.~\cite{PhysRevLett.118.207701}.

We now return to discuss the subtraction of the logarithmic UV divergence in  the integral Eq.~\eqref{eq:finintpol}. Note that the second, convergent, term has support only close to the Fermi surface (i.e., at $k\approx 0$) while the first, UV divergent part gets contributions from the full Fermi sea. Since anomalies are expressed in IR phenomena, it is plausible to subtract the first term and keep the second. This was done in Ref.~\cite{PhysRevLett.118.207701}, where it was shown that it gives results consistent with a numerical simulation of the full Hamiltonian Eq.~\eqref{parent}. It is reassuring that these various methods give the same result for the current, and satisfying that our derivations make a close connection to the chiral anomaly.

\section{DETAILS OF SPHERE AND CYLINDER SOLUTIONS} \label{app:sol}
\subsection{Sphere}
Following Eqs.~\eqref{alpha_equation} and \eqref{square_root}, we define $\alpha$ and $\sqrt\alpha$:
\begin{eqnarray}
\alpha_1 ~&=&~ \gamma \, + \, \frac{\beta}{2} \bigl( -\beta \, + \, i \sqrt{4\gamma - \beta^2} \,  \bigr), \quad \\
\sqrt \alpha_1 ~&=&~ \sqrt{\gamma - \frac{\beta^2}{4} } + i \frac{\beta}{2} = p + iq 
\end{eqnarray} 
The reference solution has the form
\begin{eqnarray}
B_r &=&\frac{2}{\alpha r^3} \left[\sinh \left(\sqrt \alpha r\right) - \sqrt \alpha r \cosh\left(\sqrt \alpha r\right)\right] \cos\theta ,\\
B_\theta &=& \frac{1}{\alpha r^3} \left[\left(1 + \alpha r^2\right) \sinh\left(\sqrt \alpha r\right) - \sqrt \alpha \
r \cosh\left(\sqrt \alpha r\right) \right] \sin\theta,\\
B_\phi &=& \frac{1}
    {r^2} \left[\sinh\left(\sqrt \alpha r\right) - \sqrt \alpha r \cosh\left(\sqrt \alpha r\right)\right] \sin\theta 
\end{eqnarray} 
and since we must take $\alpha$ and $\sqrt \alpha$ 
complex, we get complex fields, whose real and imaginary parts satisfy our Eq.\,(\ref{generalized_London}) separately, that we must combine to satisfy the boundary conditions.

The real part $B_r$, after considerable algebra, reads
\begin{eqnarray}
B^r_r ~&=&~ \frac{\cos \theta}{\gamma^2 r^3}  \nonumber \\
&\times& \sqrt{4 \gamma -\beta ^2} \cosh [\frac{1}{2} r \sqrt{4 \gamma -\beta ^2}] \left(\beta  \sin \frac{\beta  r}{2} - \gamma  r \cos \frac{\beta  r}{2}\right) -\sinh [\frac{1}{2} r \sqrt{4 \gamma -\beta ^2} ]\left(\left(\beta^2-2 \gamma \right) \cos \frac{\beta  r}{2} + \beta  \gamma  r \sin \frac{\beta  r}{2} \right) \nonumber \\
B^r_\theta ~&=&~ \frac{\sin \theta} {2 \gamma^2 r^3}  \nonumber \\
 &\times& \sqrt{4 \gamma -\beta ^2} \cosh [\frac{1}{2} r \sqrt{4 \gamma -\beta ^2}] \left(\beta  \sin \frac{\beta  r}{2}-\gamma  r \cos \frac{\beta  r}{2}\right)-\sinh [\frac{1}{2} r \sqrt{4 \gamma -\beta ^2}] \left(\left(\beta ^2-2 \gamma  \left(\gamma  r^2+1\right)\right) \cos \frac{\beta  r}{2}+\beta  \gamma  r \sin \frac{\beta  r}{2}\right) \nonumber \\
B^r_\phi ~&=&~ \frac{\sin \theta}{2 \gamma  r^2}  \nonumber \\
 &\times& \sinh [\frac{1}{2} r \sqrt{4 \gamma -\beta ^2}] \left(2 \gamma  r \sin \frac{\beta  r}{2}+\beta  \cos \frac{\beta  r}{2}\right)-\sqrt{4 \gamma -\beta ^2} \sin \frac{\beta  r}{2} \cosh [\frac{1}{2} r \sqrt{4 \gamma -\beta ^2}]
\end{eqnarray}
and the imaginary part $B^i$ reads
\begin{eqnarray}
B^i_r ~&=&~ \frac{\cos \theta}{\gamma ^2 r^3} \nonumber \\
 &\times&\cosh \left[\frac{1}{2} r \sqrt{4 \gamma -\beta ^2}\right] \left(\beta  \gamma  r \cos \frac{\beta  r}{2}-\left(\beta ^2-2 \gamma \right) \sin \frac{\beta  r}{2} \right)-\sqrt{4 \gamma -\beta ^2} \sinh \left[\frac{1}{2} r \sqrt{4 \gamma -\beta ^2}\right] \left(\gamma  r \sin \frac{\beta  r}{2}+\beta  \cos \frac{\beta  r}{2}\right) \nonumber  \\
B^i_\theta ~&=&~ \frac{\sin \theta} {2 \gamma^2 r^3}  \nonumber \\ 
 &\times& \cosh \left[ \frac{1}{2} r \sqrt{4 \gamma -\beta^2} \right] 
\left( 2 \gamma  \bigl( \left( \gamma  r^2+1\right)-\beta ^2 \right) \sin \frac{\beta  r}{2}  
+ \beta  \gamma  r \cos \frac{\beta  r}{2} \bigr) \nonumber \\
 &-& \sqrt{4 \gamma -\beta ^2} \sinh \left[ \frac{1}{2} r \sqrt{4 \gamma -\beta^2} \right] \left(\gamma  r \sin \frac{\beta  r}{2}+\beta  \cos \frac{\beta  r}{2} \right) \nonumber \\
B^i_\phi ~&=&~ \frac{\sin \theta }{2 \gamma  r^2}  \nonumber \\
 &\times& \sqrt{4 \gamma -\beta ^2} \cos \frac{\beta  r}{2} \sinh \left[\frac{1}{2} r \sqrt{4 \gamma -\beta ^2} \right]+\cosh \left[\frac{1}{2} r \sqrt{4 \gamma -\beta ^2}\right] \left(\beta  \sin \frac{\beta  r}{2} - 2 \gamma  r \cos \frac{\beta  r}{2}\right)
\end{eqnarray}

We note that this is the solution inside the superconducting sphere, and that it is thus valid only if $r\leq R$. To satisfy the boundary condition $j_r (R) = 0$, we must take a linear superposition $B^r + \eta B^i$ to get the full solution inside the superconducting sphere.  One finds
\begin{align}
\eta = \frac {\sinh [\frac{1}{2} R \sqrt{4 \gamma -\beta ^2}] \left(2 \gamma  R \sin \frac{\beta  R}{2} +\beta  \cos \frac{\beta  R}{2}\right) -\sqrt{4 \gamma -\beta ^2} \sin \frac{\beta  R}{2} \cosh [\frac{1}{2} R \sqrt{4 \gamma -\beta ^2}] }{\cosh [\frac{1}{2} R \sqrt{4 \gamma -\beta ^2}] \left(2 \gamma  R \cos \frac{\beta  R}{2}-\beta  \sin \frac{\beta  R}{2}\right)-\sqrt{4 \gamma -\beta ^2} \cos \frac{\beta  R}{2} \sinh [\frac{1}{2} R \sqrt{4 \gamma -\beta ^2}]}.
\end{align}
Thus, the final solution for the magnetic field inside the superconducting sphere reads
\begin{equation}
\vec{B}^{\text{in}}_{\text{sphere}} = \left(B_r^r,B^r_{\theta},B^r_{\phi}\right) + \eta \left(B^i_r,B^i_{\theta},B^i_{\phi}\right)
\end{equation}

Outside the sphere, i.e., for $r>R$, the magnetic field takes the usual dipole expansion form according to the solution of London, and reads \cite{London1950},
\begin{align}
B^{\text{out}}_r(r,\theta) &= \left(H_0+\frac{2M}{r^3}\right) \cos{\theta},
\\
B^{\text{out}}_{\theta}(r,\theta)&= \left(-H_0+\frac{M}{r^3}\right)\sin{\theta},
\end{align}
with
\begin{align}
M &= \frac{R^3}{3}\left[\frac{B^{\text{in}}_r(R,\theta)}{\cos{\theta}}+\frac{B^{\text{in}}_{\theta}(R,\theta)}{\sin{\theta}}\right],
\\
H_0 &= \frac{1}{3}\left[\frac{B^{\text{in}}_r(R,\theta)}{\cos{\theta}}-2\frac{B^{\text{in}}_{\theta}(R,\theta)}{\sin{\theta}}\right].
\end{align}

From the internal solutions, one can calculate the corresponding currents $\vec{j}_{\text{sphere}} = \vec{\nabla}\times \vec{B}^{\text{in}}_{\text{sphere}}$, which can be used to explicitly calculate the magnetic anapole moment, which is given by
\begin{equation}
    T_i = \frac{1}{10c}\int\left[r_i\left(\vec{r}\cdot \vec{j}\right)-2r^2J_i\right] d^3x.
\end{equation}
In the present situation, $T_x$ and $T_y$ are both zero, but $T_z$ takes a finite value. On the unit sphere, and in units where $c=1$, it explicitly reads,
\begin{equation}
    T_z = -2\pi\frac{-6\beta\left(1+\gamma\right)\cosh{\left(2\delta^{-1}\right)} + 2\left[-3\beta\left(\gamma-1\right)\cos{\beta} + \left(3\beta^2-\gamma^2\right)\sin{\beta}\right]+\beta\delta\left[-3\beta^2+\gamma\left(12+\gamma\right)\right]\sinh{\left(2\delta^{-1}\right)}}{3\gamma^2\left[\delta\cosh{\left(\delta^{-1}\right)}\left(2\gamma \cos{\frac{\beta}{2}}-\beta\sin{\frac{\beta}{2}}\right)-2\cos{\frac{\beta}{2}}\sinh{\left(\delta^{-1}\right)}\right]},
\end{equation}
where $\delta = \frac{2}{\sqrt{4\gamma-\beta^2}}$ is the penetration depth.

\subsection{Cylinder}

In the case of a cylinder, we take real and imaginary parts of an ansatz consisting of modified Bessel functions of the second kind:
$$B_r = 0, \quad B_{\phi} =\sqrt \alpha \kappa K_1(r \sqrt \alpha), \quad B_z = K_0(r\sqrt \alpha)$$.
As in the spherical case, we take the real and imaginary parts independently and find coefficients to satisfy boundary conditions $B_z = B_0, \quad B_\phi = 0$, fixing $R=1$. Since this brings in Bessel functions, it must be done numerically for specific values $\beta$ and $\gamma$.

We looked at a few cases to get numerical solutions and here are the answers that we get for the azimuthal components of the magnetic field.

For $\beta = 15, \gamma = 57$, we find 
\begin{eqnarray}
&{}&
( -4.77814 B_0)\, {\rm Re} \left\{
  K_1\left[\left(\frac{15 i}{2}+\frac{\sqrt{3}}{2}\right) r\right]\right\}
 \nonumber \\ 
&+&(-2.14645  B_0)\,  {\rm Im\/} \left\{
 K_1\left[
 \left(\frac{15 i}{2}+ \frac{\sqrt{3}}{2}\right) r
 \right] 
 \right\}.
\end{eqnarray}

 For $\beta = 10, \gamma = 4$, we find
\begin{eqnarray}
&{}&
( -69.1343 B_0)\, {\rm Re} \left\{
  K_1\left[\left(5i +\sqrt{15}\right) r\right]\right\}
 \nonumber \\ 
&+&(69.2327 B_0)\,  {\rm Im\/} \left\{
  K_1\left[\left(5i +\sqrt{15}\right) r\right]\right\}.
\end{eqnarray}

For $\beta= 1, \gamma = 200$, we find
\begin{eqnarray}
&{}&
( 2.06099 \times 10^6 B_0)\, {\rm Re} \left\{
  K_1\left[\left(\frac{i}{2}+\frac{\sqrt{799}}{2}\right) r\right]\right\} \nonumber \\ 
&+&(3.61166 \times 10^6 B_0)\,  {\rm Im\/} \left\{
  K_1\left[\left(\frac{i}{2}+\frac{\sqrt{799}}{2}\right) r\right]\right\}.
\end{eqnarray}

\end{widetext}


\bibliography{AdotB}
\end{document}